\documentclass{article}
\usepackage{spconf,amsmath,graphicx}

\usepackage{subcaption} 
\usepackage{amsmath}
\usepackage{bm} 
\usepackage{amsfonts}
\usepackage{mathtools} 
\usepackage{multirow} 
\usepackage[dvipsnames]{xcolor} 
\usepackage{pgfplots}
\usepackage{stfloats} 
\usepackage{url}
\usepackage{hyperref}
\usepackage{booktabs}
\usepackage{soul} 
\usepackage{caption} 

\pgfplotsset{compat=newest}
\usetikzlibrary{
    patterns,
}
\usepgfplotslibrary{groupplots}
\newcommand{\etal}{\emph{et al.}\ }
\newcommand{\RomanNumeralCaps}[1]{\MakeUppercase{\romannumeral #1}}

\usepackage{circledsteps}
\newcommand{\numbercircled}[1]{\raisebox{.5pt}{\textcircled{\raisebox{-.9pt} {#1}}}}

\title{Defending Your Voice: Adversarial Attack on Voice Conversion}
\name{
    Chien-yu Huang$^*$,
    Yist Y. Lin$^*$,
    Hung-yi Lee,
    Lin-shan Lee
    \thanks{$^*$ These authors contributed equally.}
}
\address{College of Electrical Engineering and Computer Science, National Taiwan University, Taiwan}

\begin{document}
%
\maketitle
\begin{abstract}
    Substantial improvements have been achieved in recent years in voice conversion, which converts the speaker characteristics of an utterance into those of another speaker without changing the linguistic content of the utterance.
    Nonetheless, the improved conversion technologies also led to concerns about privacy and authentication.
    It thus becomes highly desired to be able to prevent one's voice from being improperly utilized with such voice conversion technologies.
    This is why we report in this paper the first known attempt to perform adversarial attack on voice conversion.
    We introduce human imperceptible noise into the utterances of a speaker whose voice is to be defended.
    Given these adversarial examples, voice conversion models cannot convert other utterances so as to sound like being produced by the defended speaker.
    Preliminary experiments were conducted on two currently state-of-the-art zero-shot voice conversion models.
    Objective and subjective evaluation results in both white-box and black-box scenarios are reported.
    It was shown that the speaker characteristics of the converted utterances were made obviously different from those of the defended speaker, while the adversarial examples of the defended speaker are not distinguishable from the authentic utterances.
\end{abstract}
\begin{keywords}
    voice conversion, adversarial attack, speaker verification, speaker representation
\end{keywords}
\section{Introduction}
\label{sec:intro}

Voice conversion aims to alter some specific acoustic characteristics of an utterance, such as the speaker identity, while preserving the linguistic content.
These technologies were made much more powerful by deep learning \cite{Chou2018, CycleGAN-VC2, kameoka2018stargan, Chou2019, pmlr-v97-qian19c}, but the improved technologies also led to concerns about privacy and authentication.
One's identity may be counterfeited by voice conversion and exploited in improper ways, which is only one of the many deepfake problems observed today generated by deep learning, such as synthesized fake photos or fake voice.
Detecting any of such artifacts or defending against such activities is thus increasingly important \cite{Sahidullah2015, dolhansky2020deepfake, Li_2019_CVPR_Workshops, FakeCatcher}, which applies equally to voice conversion.

On the other hand, it has been widely known that neural networks are fragile in the presence of some specific noise, or prone to yield different or incorrect results if the input is disturbed by such subtle perturbations imperceptible to humans \cite{szegedy2013intriguing}.
Adversarial attack is to generate such subtle perturbations that can fool the neural networks.
It has been successful on some discriminative models \cite{goodfellow2014explaining, kurakin2016adversarial, madry2018towards}, but less reported on generative models \cite{kos2018adversarial}.

In this paper, we propose to perform adversarial attack on voice conversion to prevent one's speaker characteristics from being improperly utilized with voice conversion.
Human-imperceptible perturbations are added to the utterances produced by the speaker to be defended.
Three different approaches, the end-to-end attack, embedding attack, and feedback attack are proposed, such that the speaker characteristics of the converted utterances would be made very different from those of the defended speaker.
We conducted objective and subjective evaluations on two recent state-of-the-art zero-shot voice conversion models.
Objective speaker verification showed the converted utterances were significantly different from those produced by the defended speaker, which was then verified by subjective similarity test.
The effectiveness of the proposed approaches was also verified for black-box attack via a proxy model closer to the real application scenario.

\section{Related works}

\subsection{Voice conversion}

Traditionally, parallel data are required for voice conversion, or the training utterances of the two speakers must be paired and aligned.
To overcome this problem, Chou \etal \cite{Chou2018} obtained disentangled representations respectively for linguistic content and speaker information with adversarial training;
CycleGAN-VC \cite{CycleGAN-VC2} used cycle-consistency to ensure the converted speech to be linguistically meaningful with the target speaker's features;
and StarGAN-VC \cite{kameoka2018stargan} introduced conditional input for many-to-many voice conversion.
All these are limited to speakers seen in training.

Zero-shot approaches then tried to convert utterances to any speaker given only one example utterance without fine-tuning, and the target speaker is not necessarily seen before.
Chou \etal \cite{Chou2019} employed adaptive instance normalization for this purpose;
\textsc{AutoVC} \cite{pmlr-v97-qian19c} integrated a pre-trained d-vector and an encoder bottleneck, achieving the state-of-the-art results.

\subsection{Attacking and defending voice}

Automatic speech recognition (ASR) systems have been shown to be prone to adversarial attacks.
Applying perturbations on the waveforms, spectrograms, or MFCC features was able to make ASR systems fail to recognize the speech correctly \cite{carlini2018audio, Schoenherr2019, Alzantot2017, Taori2019, NIPS2017_7273}.
Similar goals were achieved on speaker recognition by generating adversarial examples to fool automatic speaker verification (ASV) systems to predict that these examples had been uttered by a specific speaker \cite{liu2019adversarial, li2019adversarial, Kreuk2018}.
Different approaches for spoofing ASV were also proposed to show the vulnerabilities of such systems \cite{lau2004, wu2012, Evans2013}.
But to our knowledge, applying adversarial attacks on voice conversion has not been reported yet.

On the other hand, many approaches were proposed to defend one's voice when ASV systems were shown to be vulnerable to spoofing attacks \cite{qian2016, Lavrentyeva2017, Valenti2018, chen2017you}.
In addition to the ASVspoof challenges for spoofing techniques and countermeasures \cite{Todisco2019}, Liu \etal \cite{liu2019adversarial} conducted adversarial attacks on those countermeasures, showing the fragility of them.
Obviously, all neural network models are under the threat of adversarial attacks \cite{goodfellow2014explaining}, which led to the idea of attacking voice conversion models as proposed here.

\section{Methodologies}
\label{sec:method}

\begin{figure*}[t]
    \centering
    \includegraphics[width=.85\linewidth]{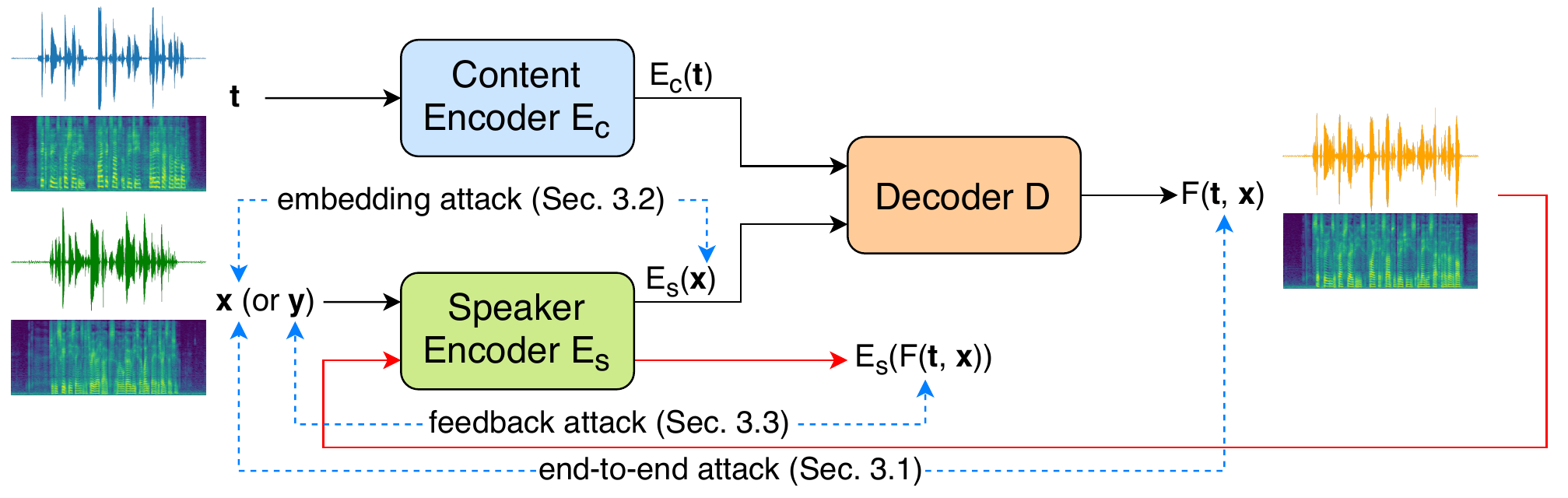}
    \caption{
        The encoder-decoder based voice conversion model and the three proposed approaches.
        Perturbations are updated on the utterances providing speaker characteristics, as the blue dashed lines indicate.
    }
    \label{fig:diagram_of_methodologies}
\end{figure*}

A widely used model for voice conversion adopted an encoder-decoder structure, in which the encoder is further divided into a content encoder and a speaker encoder, as shown in Fig.~\ref{fig:diagram_of_methodologies}.
This paper is also based on this model.
The content encoder $E_c$ extracts the content information from an input utterance $\bm{t}$ yielding $E_c(\bm{t})$, while the speaker encoder $E_s$ embeds the speaker characteristics of an input utterance $\bm{x}$ as a latent vector $E_s(\bm{x})$, as in the left part of Fig.~\ref{fig:diagram_of_methodologies}.
Taking $E_c(\bm{t})$ and $E_s(\bm{x})$ as the input, the decoder $D$ generates a spectrogram $F(\bm{t}, \bm{x})$ with content information based on $E_c(\bm{t})$ and speaker characteristics based on $E_s(\bm{x})$.

Here we only focus on the utterances fed into the speaker encoder, since we are defending the speaker characteristics provided by such utterances.
Motivated by the prior work \cite{kos2018adversarial}, here we present three approaches to performing the attack, with the target being either the output spectrogram $F(\bm{t}, \bm{x})$ (Sec.~\ref{sec:e2e_attack}), or the speaker embedding $E_s(\bm{x})$ (Sec.~\ref{sec:emb_attack}), or the combination of the two (Sec.~\ref{sec:feedback_attack}), as shown in Fig.~\ref{fig:diagram_of_methodologies}.

\subsection{End-to-end attack}
\label{sec:e2e_attack}

A straight-forward approach to perform adversarial attack on the above model in Fig.~\ref{fig:diagram_of_methodologies} is to take the decoder output $F(\bm{t}, \bm{x})$ as the target, referred to as end-to-end attack also shown in Fig.~\ref{fig:diagram_of_methodologies}.
Denote the original spectrogram of an utterance produced by the speaker to be defended as $\bm{x} \in \mathbb{R}^{M \times T}$ and the adversarial perturbation on $\bm{x}$ as $\bm{\delta} \in \mathbb{R}^{M \times T}$, where $M$ and $T$ are the total number of frequency components and time frames respectively.
An untargeted attack simply aims to alter the output of the voice conversion model and can be expressed as:
\begin{equation}
    \begin{aligned}
        & \underset{\bm{\delta}}{\text{maximize}} & & \mathcal{L}(F(\bm{t}, \bm{x} + \bm{\delta}), F(\bm{t}, \bm{x})) \\
        & \text{subject to} & & \|\bm{\delta}\|_{\infty} < \epsilon
    \end{aligned}
\end{equation}
$\mathcal{L}(\cdot, \cdot)$ is the distance between two vectors or the spectrograms for two signals and $\epsilon$ is a constraint making the perturbation subtle.
The signal $\bm{t}$ can be arbitrary offering the content of the output utterance, on which we do not focus here.

Given a certain utterance $\bm{y}$ produced by a target speaker, we can formulate a targeted attack for output signal with specific speaker characteristics:
\begin{equation}
    \label{eq:targeted_attack}
    \begin{aligned}
        & \underset{\bm{\delta}}{\text{minimize}} &  &
        \begin{multlined}
            \mathcal{L}(F(\bm{t}, \bm{x} + \bm{\delta}), F(\bm{t}, \bm{y})) \\
            - \lambda\mathcal{L}(F(\bm{t}, \bm{x} + \bm{\delta}), F(\bm{t}, \bm{x}))
        \end{multlined} \\
        & \text{subject to} &  & \|\bm{\delta}\|_{\infty} < \epsilon
    \end{aligned}
\end{equation}
The first term in the first expression in (\ref{eq:targeted_attack}) aims to make the model output sound like being produced by the speaker of $\bm{y}$, while the second term is to eliminate the original speaker identity in $\bm{x}$.
$\lambda$ is a positive valued hyperparameter balancing the importance between source and target.

To effectively constrain the range of perturbation within $[-\epsilon, \epsilon]$ while solving (\ref{eq:targeted_attack}), we adopt the approach of \emph{Change of variable} as was done previously \cite{Carlini2017} using $\tanh(\cdot)$ function.
In this way (\ref{eq:targeted_attack}) above becomes (\ref{eq:targeted_attack_with_tanh}) below:
\begin{equation}
    \label{eq:targeted_attack_with_tanh}
    \begin{aligned}
        & \underset{\bm{w}}{\text{minimize}} &  &
        \begin{multlined}
            \mathcal{L}(F(\bm{t}, \bm{x} + \bm{\delta}), F(\bm{t}, \bm{y})) \\
            - \lambda\mathcal{L}(F(\bm{t}, \bm{x} + \bm{\delta}), F(\bm{t}, \bm{x}))
        \end{multlined} \\
        & \text{subject to} &  & \bm{\delta} = \epsilon \cdot \tanh ( \bm{w} )
    \end{aligned}
\end{equation}
where $\bm{w} \in \mathbb{R}^{M \times T}$.
The clipping function is not needed here.

\subsection{Embedding attack}
\label{sec:emb_attack}

The speaker encoder $E_s$ in Fig.~\ref{fig:diagram_of_methodologies} embeds an utterance into a latent vector.
These latent vectors for utterances produced by the same speaker tend to cluster closely together, while those by different speakers tend to be separated apart.
The second approach proposed here is focused on the speaker encoder by directly changing the speaker embeddings of the utterances, referred to as embedding attack also in Fig.~\ref{fig:diagram_of_methodologies}.
As the decoder $D$ produces the output $F(\bm{t}, \bm{x})$ with speaker characteristics based on the speaker embedding $E_s(\bm{x})$ as in Fig.~\ref{fig:diagram_of_methodologies}, changing the speaker embeddings thus alters the output of the decoder.

Following the notations and expressions in (\ref{eq:targeted_attack_with_tanh}), we have:
\begin{equation}
    \label{eq:emb_attack}
    \begin{aligned}
        & \underset{\bm{w}}{\text{minimize}} &  &
        \begin{multlined}
            \mathcal{L}(E_s(\bm{x} + \bm{\delta}), E_s(\bm{y})) \\
            - \lambda\mathcal{L}(E_s(\bm{x} + \bm{\delta}), E_s(\bm{x})))
        \end{multlined} \\
        & \text{subject to} &  & \bm{\delta} = \epsilon \cdot \tanh ( \bm{w} )
    \end{aligned}
\end{equation}
where the adversarial attack is now performed with the speaker encoder $E_s$ only.
Since only the speaker encoder is involved, it is therefore more efficient.

\subsection{Feedback attack}
\label{sec:feedback_attack}

The third approach proposed here tries to combine the above two approaches by feeding the output spectrogram $F(\bm{t}, \bm{x} + \bm{\delta})$ from the decoder $D$ back to the speaker encoder $E_s$ (the red feedback loop in Fig.~\ref{fig:diagram_of_methodologies}), and consider the speaker embedding obtained in this way.
More specifically, $E_s(\bm{x} + \bm{\delta})$ in (\ref{eq:emb_attack}) is replaced by $E_s(F(\bm{t}, \bm{x} + \bm{\delta}))$ in (\ref{eq:feedback_attack}).
This is referred to as the feedback attack also in Fig.~\ref{fig:diagram_of_methodologies}.
\begin{equation}
    \label{eq:feedback_attack}
    \begin{aligned}
        & \underset{\bm{w}}{\text{minimize}} &  &
        \begin{multlined}
            \mathcal{L}(E_s(F(\bm{t}, \bm{x} + \bm{\delta})), E_s(\bm{y})) \\
            - \lambda\mathcal{L}(E_s(F(\bm{t}, \bm{x} + \bm{\delta})), E_s(\bm{x})))
        \end{multlined} \\
        & \text{subject to} &  & \bm{\delta} = \epsilon \cdot \tanh ( \bm{w} )
    \end{aligned}
\end{equation}

\section{Experimental settings}

We conducted experiments on the model proposed by Chou \etal \cite{Chou2019} (referred to as Chou's model below) and \textsc{AutoVC}.
Both were able to perform zero-shot voice conversion on unseen speakers given their few utterances without fine-tuning, considered suitable for our scenarios.
Models such as StarGAN-VC, on the other hand, are limited to the voice produced by speakers seen during training, which makes it less likely to counterfeit other speakers' voices, and we thus do not consider them here.

\subsection{Speaker encoders}

For Chou's model, all modules were trained jointly from scratch on the CSTR VCTK Corpus \cite{veaux2016vctk}.
The speaker encoder took a 512-dim mel spectrogram and generated a 128-dim speaker embedding.
\textsc{AutoVC} utilized pre-trained d-vector \cite{wan2018generalized} as speaker encoder, with 80-dim mel spectrogram as input and 256-dim speaker embedding as output, pre-trained on VoxCeleb1 \cite{Nagrani2017} and LibriSpeech \cite{panayotov2015librispeech} but generalizable to unseen speakers.

\subsection{Vocoders}

In inference, Chou's model leveraged Griffin-Lim algorithm \cite{1164317} to synthesize the audio.
\textsc{AutoVC} previously adopted WaveNet \cite{van_den_Oord+2016} as the spectrogram inverter, but here we used WaveRNN-based vocoder \cite{Lorenzo-Trueba2019} pre-trained on the VCTK corpus to generate waveforms with similar quality due to time limitation.

As we introduced perturbation on spectrogram, vocoders converting the spectrogram into waveform were necessary.
We respectively adopted Griffin-lim algorithm and WaveRNN-based vocoder for attacks on Chou's model and \textsc{AutoVC}.

\subsection{Attack scenarios}

Two scenarios were tested here.
In the first scenario, the attacker has full access to the model to be attacked.
With the complete architecture plus all trained parameters of the model being available, we can apply adversarial attack directly.
This scenario is referred to as \textbf{white-box} scenario, in which all experiments were conducted on Chou's model and \textsc{AutoVC} with their publicly available network parameters and evaluated on exactly the same models.

The second one is referred to as \textbf{black-box} scenario, in which the attacker cannot directly access the parameters of the model to be attacked, or the architecture might even be unknown.
For attacking Chou's model, we trained a new model with the same architecture but different initialization, whereas for \textsc{AutoVC} we trained a new speaker encoder with the architecture similar to the one in the original \textsc{AutoVC}.
These newly-trained models are then used as proxy models to generate adversarial examples to be evaluated with the publicly available ones in the same way as in white-box scenario.

\subsection{Attack procedure}

We selected $\ell_2$ norm as $\mathcal{L}(\cdot, \cdot)$, and $\lambda = 0.1$ for all experiments.
$\bm{w}$ in (\ref{eq:targeted_attack_with_tanh}),~(\ref{eq:emb_attack}),~(\ref{eq:feedback_attack}) was initialized from a standard normal distribution.
The perturbations to be added to the utterances was $\epsilon\tanh (\bm{w})$.
Adam \cite{DBLP:journals/corr/KingmaB14} optimizer was adopted to update $\bm{w}$ iteratively according to the loss function defined in (\ref{eq:targeted_attack_with_tanh}), (\ref{eq:emb_attack}), (\ref{eq:feedback_attack}) with the learning rate being 0.001 and the number of iterations being 1500.

\section{Results}

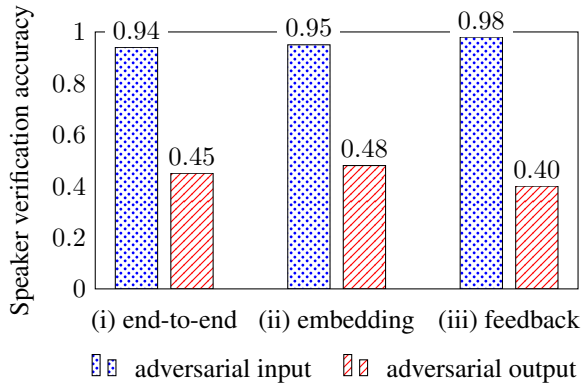
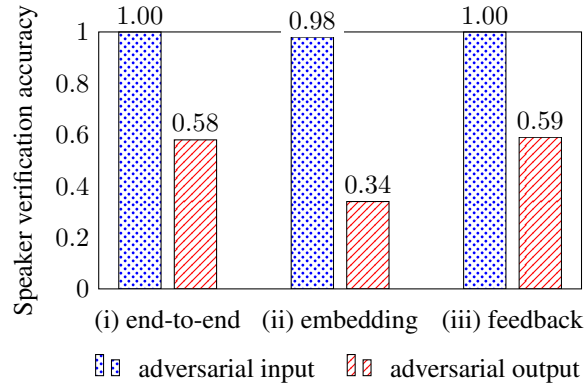
\begin{figure*}[t]
    \begin{subfigure}{.5\linewidth}
        \centering
        \begin{tikzpicture}
            \begin{axis}[
                    height=5cm,
                    width=8cm,
                    ybar=5pt,
                    ylabel={Speaker verification accuracy},
                    ymin=0,
                    ymax=1,
                    symbolic x coords={
                            {(i) end-to-end},
                            {(ii) embedding},
                            {(iii) feedback}
                        },
                    xtick=data,
                    xtick style={
                            draw=none
                        },
                    enlarge x limits=0.2,
                    legend style={
                            draw=none,
                            at={(0.5,-0.2)},
                            anchor=north,
                            legend columns=-1,
                            inner sep=5pt,
                            column sep=5pt,
                            /tikz/every even column/.append style={column sep=10pt}
                        },
                    bar width=16pt,
                    nodes near coords,
                    nodes near coords align={vertical},
                    nodes near coords style={
                            fill=white,
                            inner sep=3pt,
                            /pgf/number format/.cd,
                            fixed,
                            fixed zerofill,
                            precision=2,
                        },
                    ytick style={
                            draw=none,
                        },
                ]
                \addplot[
                    ybar,
                    pattern=crosshatch dots,
                    pattern color=blue,
                ] coordinates {
                        ({(i) end-to-end}, 0.94)
                        ({(ii) embedding}, 0.95)
                        ({(iii) feedback}, 0.98)
                    };
                \addplot[
                    ybar,
                    pattern=north east lines,
                    pattern color=red,
                ] coordinates {
                        ({(i) end-to-end}, 0.45)
                        ({(ii) embedding}, 0.48)
                        ({(iii) feedback}, 0.4)
                    };

                \legend{
                    adversarial input,
                    adversarial output
                }
            \end{axis}
        \end{tikzpicture}
        \caption{Chou's}
        \label{fig:one_shot_white_box}
    \end{subfigure}%
    \begin{subfigure}{.5\linewidth}
        \centering
        \begin{tikzpicture}
            \begin{axis}[
                    height=5cm,
                    width=8cm,
                    ybar=5pt,
                    ylabel={Speaker verification accuracy},
                    ymin=0,
                    ymax=1,
                    symbolic x coords={
                            {(i) end-to-end},
                            {(ii) embedding},
                            {(iii) feedback}
                        },
                    xtick=data,
                    xtick style={
                            draw=none
                        },
                    enlarge x limits=0.2,
                    legend style={
                            draw=none,
                            at={(0.5,-0.2)},
                            anchor=north,
                            legend columns=-1,
                            inner sep=5pt,
                            column sep=5pt,
                            /tikz/every even column/.append style={column sep=10pt}
                        },
                    bar width=16pt,
                    nodes near coords,
                    nodes near coords align={vertical},
                    nodes near coords style={
                            fill=white,
                            inner sep=3pt,
                            /pgf/number format/.cd,
                            fixed,
                            fixed zerofill,
                            precision=2,
                        },
                    ytick style={
                            draw=none,
                        },
                ]
                \addplot[
                    ybar,
                    pattern=crosshatch dots,
                    pattern color=blue,
                ] coordinates {
                        ({(i) end-to-end}, 1.00)
                        ({(ii) embedding}, 0.98)
                        ({(iii) feedback}, 1.00)
                    };
                \addplot[
                    ybar,
                    pattern=north east lines,
                    pattern color=red,
                ] coordinates {
                        ({(i) end-to-end}, 0.58)
                        ({(ii) embedding}, 0.34)
                        ({(iii) feedback}, 0.59)
                    };

                \legend{
                    adversarial input,
                    adversarial output
                }
            \end{axis}
        \end{tikzpicture}
        \caption{\textsc{AutoVC}}
        \label{fig:autovc_white_box}
    \end{subfigure}
    \caption{
        Speaker verification accuracy for the defended speaker with \emph{Chou's} ($\epsilon = 0.075$) and \textsc{AutoVC} ($\epsilon = 0.05$) by the three proposed approaches under the white-box scenario.
    }
    \label{fig:compare_methodologies_white_box}
\end{figure*}

\subsection{Objective tests}

For automatic evaluation, we adopted speaker verification accuracy as a reliable metric.
The speaker verification system used here first encoded two input utterances into embeddings and then computed the similarity between the two.
The two utterances were considered to be uttered by the same speaker if the similarity exceeds a threshold.
In the test, each time we compared a machine-generated utterance with a real utterance providing the speaker characteristics in the experiments ($\bm{x}$ in Fig.~\ref{fig:diagram_of_methodologies}), and the \textbf{speaker verification accuracy} used below is defined as the percentage of the cases that the two are considered to be produced by the same speaker by the speaker verification system.

The verification system used here was based on a pre-trained d-vector model \footnote{\url{https://github.com/resemble-ai/Resemblyzer}} which was different from the speaker encoders of the two attacked models.
The threshold was determined based on the equal error rate (EER) when verifying utterance pairs randomly sampled from the VCTK corpus in the following way.
We sampled 256 utterances for each speaker in the dataset, half of which were used as positive samples and the other half as negative ones.
For positive samples, the similarity was computed with random utterances of the authentic speaker, whereas the negative ones were computed against random utterances produced by other randomly selected speakers.
This gave the threshold of 0.683 with the EER being 0.056.

We randomly collected enough number of utterances offering the speaker characteristics ($\bm{x}$ in Fig.~\ref{fig:diagram_of_methodologies}) from 109 speakers in the VCTK corpus and enough number of utterances offering the content ($\bm{t}$ in Fig.~\ref{fig:diagram_of_methodologies}), and performed voice conversion with both Chou's and \textsc{AutoVC} to generate the \textbf{original outputs} ($F(\bm{t}, \bm{x})$ in Fig.~\ref{fig:diagram_of_methodologies}).
We randomly collected 100 of such pairs ($\bm{x}, F(\bm{t}, \bm{x})$) produced by both Chou's and \textsc{AutoVC} which were considered by the speaker verification system mentioned above to be produced by the same speaker to be used for the test below.\footnote{We only evaluated the examples that can be successfully converted by the voice conversion model because if an example cannot be successfully converted, then there is no need to defend it.}
Thus the speaker verification accuracy of all these \textbf{original outputs} ($F(\bm{t}, \bm{x})$) is 1.00.
We then created corresponding adversarial examples ($\bm{x} + \bm{\delta}$ in (\ref{eq:targeted_attack_with_tanh}),~(\ref{eq:emb_attack}),~(\ref{eq:feedback_attack})), targeting speakers with gender opposite to the defended speaker, and performed speaker verification respectively on these adversarial example utterances (referred to as \textbf{adversarial input}) and the converted utterances $F(\bm{t}, \bm{x} + \bm{\delta})$ (referred to as \textbf{adversarial output}).
The same examples were used in the test for Chou's and \textsc{AutoVC}.

Fig.~\ref{fig:one_shot_white_box} shows the speaker verification accuracy for the adversarial input and output utterances evaluated with respect to the defended speaker under the white-box scenario for Chou's model.
Results for the three approaches mentioned in Sec.~\ref{sec:method} are in the three sections (i) (ii) (iii), with the blue crosshatch-dotted bar and the red diagonal-lined bar respectively for \textbf{adversarial input} and \textbf{adversarial output}.
Similarly in Fig.~\ref{fig:autovc_white_box} for \textsc{AutoVC}.
We can see the adversarial inputs sounded very close to the defended speaker or the perturbation $\bm{\delta}$ almost imperceptible (the blue bars very close to 1.00), while the converted utterances sounded as from a different speaker (the red bars much lower).
All the three approaches were effective, although feedback attack worked the best for Chou's (section (iii) in Fig.~\ref{fig:one_shot_white_box}), while embedding attack worked very well for both Chou's and \textsc{AutoVC} with respect to both adversarial input and output (section (ii) of each chart).

\begin{figure*}[t]
    \centering
    \pgfplotsset{
        every axis title/.append style={
                at={(0.45, -0.675)}
            }
    }
    \begin{tikzpicture}
        \begin{groupplot}[
                group style={group size=3 by 1},
            ]
            \nextgroupplot[
                title={(a) end-to-end},
                height=5cm,
                width=6cm,
                font=\small,
                ymin=0,
                xtick={1, 2, 3, 4, 5, 6},
                xticklabels={0.01, 0.02, 0.05, 0.075, 0.1, 0.2},
                x tick label style={
                        font=\footnotesize,
                    },
                ytick={0.0, 0.2, 0.4, 0.6, 0.8, 1.0},
                xlabel={Perturbation scale (\(\epsilon\))},
                ylabel={Speaker verification accuracy},
                every axis x label/.style={
                        at={(0.5, -0.2)},
                    },
            ]
            \addplot[color=blue, mark=*] coordinates {
                    (1, 1.00)
                    (2, 1.00)
                    (3, 1.00)
                    (4, 0.95)
                    (5, 0.70)
                    (6, 0.17)
                };
            \addplot[color=red, mark=square*] coordinates {
                    (1, 1.00)
                    (2, 1.00)
                    (3, 0.90)
                    (4, 0.63)
                    (5, 0.36)
                    (6, 0.10)
                };

            \nextgroupplot[
                title={(b) embedding},
                height=5cm,
                width=6cm,
                font=\small,
                ymin=0,
                xtick={1, 2, 3, 4, 5, 6},
                xticklabels={0.01, 0.02, 0.05, 0.075, 0.1, 0.2},
                x tick label style={
                        font=\footnotesize,
                    },
                ytick={0.0, 0.2, 0.4, 0.6, 0.8, 1.0},
                xlabel={Perturbation scale (\(\epsilon\))},
                legend style={
                        draw=none,
                        at={(0.5, -0.3)},
                        legend cell align=left,
                        anchor=north,
                        legend columns=2,
                        inner sep=0pt,
                        /tikz/every even column/.append style={column sep=5pt},
                    },
                every axis x label/.style={
                        at={(0.5, -0.2)},
                    },
            ]
            \addplot[color=blue, mark=*] coordinates {
                    (1, 1.00)
                    (2, 1.00)
                    (3, 1.00)
                    (4, 0.99)
                    (5, 0.82)
                    (6, 0.15)
                };
            \addplot[color=red, mark=square*] coordinates {
                    (1, 1.00)
                    (2, 1.00)
                    (3, 0.90)
                    (4, 0.63)
                    (5, 0.36)
                    (6, 0.1)
                };
            \legend{adversarial input, adversarial output}

            \nextgroupplot[
                title={(c) feedback},
                height=5cm,
                width=6cm,
                font=\small,
                ymin=0,
                xtick={1, 2, 3, 4, 5, 6},
                xticklabels={0.01, 0.02, 0.05, 0.075, 0.1, 0.2},
                x tick label style={
                        font=\footnotesize,
                    },
                ytick={0.0, 0.2, 0.4, 0.6, 0.8, 1.0},
                xlabel={Perturbation scale (\(\epsilon\))},
                every axis x label/.style={
                        at={(0.5, -0.2)},
                    },
            ]
            \addplot[color=blue, mark=*] coordinates {
                    (1, 1.00)
                    (2, 1.00)
                    (3, 1.00)
                    (4, 1.00)
                    (5, 0.83)
                    (6, 0.22)
                };
            \addplot[color=red, mark=square*] coordinates {
                    (1, 1.00)
                    (2, 0.99)
                    (3, 0.87)
                    (4, 0.63)
                    (5, 0.35)
                    (6, 0.11)
                };

        \end{groupplot}
    \end{tikzpicture}

    \caption{
        The speaker verification accuracy for different perturbation $\epsilon$ on \textbf{Chou's model} under \textbf{black-box} scenario for the three proposed approaches: (a) end-to-end, (b) embedding, and (c) feedback attacks.
    }
    \label{fig:compare_methodologies_black_box}
\end{figure*}
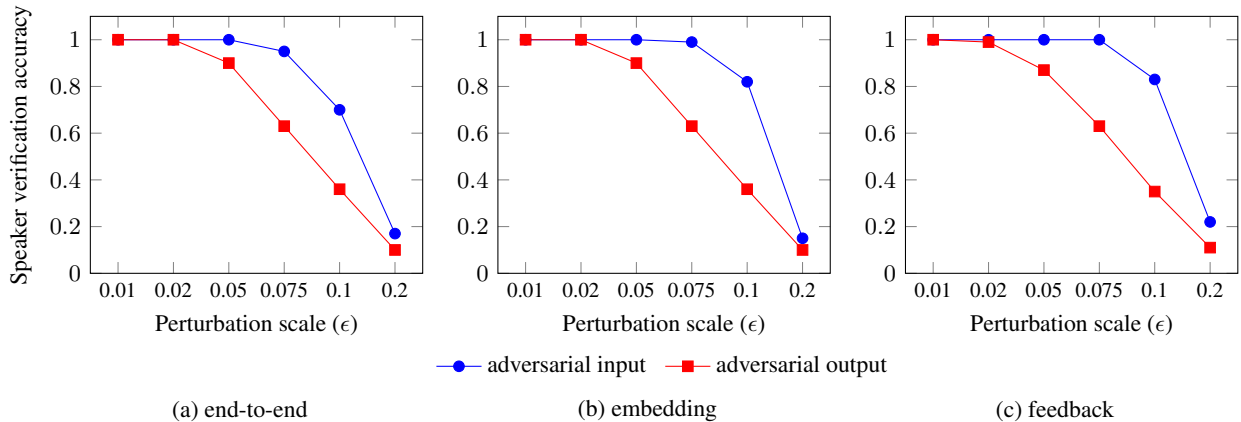

For black-box scenario, we analyzed the same speaker verification accuracy as in Fig.~\ref{fig:one_shot_white_box} for Chou's model only but with varying scale of the perturbations $\epsilon$, with results plotted in Fig.~\ref{fig:compare_methodologies_black_box} (a) (b) (c) respectively for the three approaches proposed.
We see when $\epsilon = 0.075$ the adversarial inputs were kept almost intact (blue curves close to 1.0) while adversarial outputs were seriously disturbed (red curves much lower).
However, as $\epsilon \geq 0.1$ the speaker characteristics of the adversarial inputs were altered drastically (blue curves went down), although the adversarial outputs sounded very different (red curves went very low).

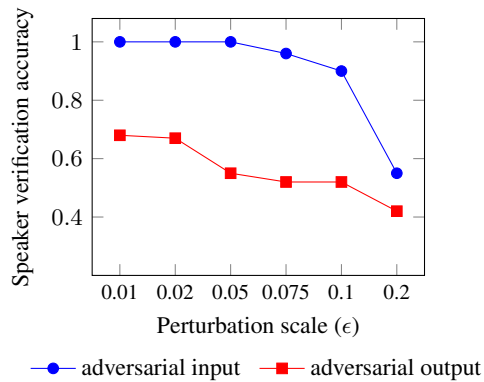
\begin{figure}[htb]
    \centering
    \begin{tikzpicture}
        \begin{axis}[
                height=5cm,
                width=6cm,
                font=\small,
                xtick={1, 2, 3, 4, 5, 6},
                xticklabels={0.01, 0.02, 0.05, 0.075, 0.1, 0.2},
                x tick label style={
                        font=\footnotesize,
                    },
                ymin=0.2,
                ytick={0.4, 0.6, 0.8, 1.0},
                xlabel={Perturbation scale (\(\epsilon\))},
                ylabel={Speaker verification accuracy},
                legend style={
                        draw=none,
                        at={(0.5, -0.3)},
                        legend cell align=left,
                        anchor=north,
                        legend columns=2,
                        inner sep=0pt,
                        /tikz/every even column/.append style={column sep=5pt},
                    },
                every axis x label/.style={
                        at={(0.5, -0.2)},
                    },
            ]
            \addplot[color=blue, mark=*] coordinates {
                    (1), 1.00)
                    (2, 1.00)
                    (3, 1.00)
                    (4, 0.96)
                    (5, 0.90)
                    (6, 0.55)
                };
            \addplot[color=red, mark=square*] coordinates {
                    (1, 0.68)
                    (2, 0.67)
                    (3, 0.55)
                    (4, 0.52)
                    (5, 0.52)
                    (6, 0.42)
                };
            \legend{adversarial input, adversarial output}
        \end{axis}
    \end{tikzpicture}
    \caption{
        Same as Fig.~\ref{fig:compare_methodologies_black_box} except on \textbf{\textsc{AutoVC}} and for \textbf{embedding attack} only.
    }
    \label{fig:autovc_black_box}
\end{figure}

Fig.~\ref{fig:autovc_black_box} shows the same results as in Fig.~\ref{fig:compare_methodologies_black_box} except on \textsc{AutoVC} with embedding attack only (as the other two methods did not work well in white-box scenario in Fig.~\ref{fig:autovc_white_box}).
We see very similar results as in Fig.~\ref{fig:compare_methodologies_black_box}, and the embedding attack worked successfully with \textsc{AutoVC} for good choices of $\bm{\epsilon}$ ($0.05 \leq \bm{\epsilon} \leq 0.075$).

Among the three proposed approaches, the embedding attack turned out to be the most attractive, considering both defending effectiveness (as mentioned above) and time efficiency.
The feedback attack offered very good performance on Chou's model, but less effective on \textsc{AutoVC}.
It also took more time to apply the perturbation as one more complete encoder-to-decoder inference was required.
It is interesting to note that the end-to-end attack offered performance comparable to the other two approaches, although it is based on the distance between spectrograms, very different from the distance between speaker embeddings, based on which the other two approaches relied on.

\subsection{Subjective tests}
\label{sec:subjective_tests}

\begin{figure*}[t]
    \centering
    \begin{tikzpicture}
        \pgfplotsset{
            every axis title/.append style={
                    at={(0.5, -0.4)}
                }
        }
        \begin{groupplot}[
                group style={group size=2 by 1},
            ]
            \nextgroupplot[
                title={(a) Chou's},
                height=8.5cm,
                width=9cm,
                ybar stacked,
                ymin=0, ymax=100,
                bar width=20pt,
                nodes near coords,
                every node near coord/.style={
                        fill=white,
                        draw=white,
                        inner sep=0.5pt,
                        font=\scriptsize
                    },
                nodes near coords style={
                        /pgf/number format/.cd,
                        fixed,
                        fixed zerofill,
                        precision=1,
                    },
                ylabel={Percentage (\%)},
                ylabel style={
                        yshift=-10mm,
                    },
                xticklabels={
                        adv. input,
                        adv. output,
                        adv. input,
                        adv. output,
                        original output
                    },
                enlarge y limits=0.0,
                enlarge x limits=true,
                xtick=data,
                xtick style={
                        draw=none
                    },
                ytick style={
                        draw=none
                    },
                x tick label style={
                        anchor=north,
                        align=center,
                        text width=1.2cm,
                        yshift=-1mm,
                        font=\linespread{0.8}\selectfont,
                    },
                extra y ticks={
                        0,
                        44.5
                    },
                extra y tick labels={
                        white-box,
                        black-box
                    },
                extra y tick style={
                        tick align=outside,
                        tick pos=left,
                        tick label style={
                                yshift=146mm,
                                xshift=-2mm,
                                rotate=0,
                            },
                        ticklabel pos=top,
                        rotate=-90,
                    },
                extra x ticks={
                        1,
                        2,
                        3.5,
                        4.5,
                        6
                    },
                extra x tick style={
                        tick align=outside,
                        tick pos=left,
                        tick label style={
                                yshift=5mm,
                            },
                        ticklabel pos=top,
                    },
                extra x tick labels={
                        \numbercircled{1},
                        \numbercircled{2},
                        \numbercircled{3},
                        \numbercircled{4},
                        \numbercircled{5}
                    },
            ]
            \node[
                text width=1em,
                anchor=north,
            ] at (rel axis cs: 0.5, 0.5) {
                \phantomsubcaption
                \label{plot:one_shot_subjective_result}
            };
            \addplot[
                ybar,
                pattern=crosshatch,
                pattern color=blue,
            ] plot coordinates {
                    (1, 0.5)
                    (2, 58.5)
                    (3.5, 0.0)
                    (4.5, 41.5)
                    (6, 1.5)
                };
            \addplot[
                ybar,
                pattern=crosshatch dots,
                pattern color=teal,
            ] plot coordinates {
                    (1, 3.0)
                    (2, 31.0)
                    (3.5, 3.5)
                    (4.5, 35.0)
                    (6, 16.0)
                };
            \addplot[
                ybar,
                pattern=north east lines,
                pattern color=orange,
            ] plot coordinates {
                    (1, 26.5)
                    (2, 10.5)
                    (3.5, 18.5)
                    (4.5, 17.5)
                    (6, 51.0)
                };
            \addplot[
                ybar,
                pattern=grid,
                pattern color=red,
            ] plot coordinates {
                    (1, 70.0)
                    (2, 0.0)
                    (3.5, 78.0)
                    (4.5, 6.0)
                    (6, 31.5)
                };

            \nextgroupplot[
                title={(b) \textsc{AutoVC}},
                height=8.5cm,
                width=9cm,
                ybar stacked,
                ymin=0, ymax=100,
                bar width=20pt,
                nodes near coords,
                every node near coord/.style={
                        fill=white,
                        draw=white,
                        inner sep=0.5pt,
                        font=\scriptsize
                    },
                nodes near coords style={
                        /pgf/number format/.cd,
                        fixed,
                        fixed zerofill,
                        precision=1,
                    },
                legend style={
                        draw=none,
                        at={(-0.1, -0.18)},
                        legend cell align=left,
                        anchor=north,
                        legend columns=4,
                        inner sep=0pt,
                        /tikz/every even column/.append style={column sep=5pt},
                        font=\small,
                    },
                xticklabels={
                        adv. input,
                        adv. output,
                        adv. input,
                        adv. output,
                        original output
                    },
                enlarge y limits=0.0,
                enlarge x limits=true,
                xtick=data,
                xtick style={
                        draw=none
                    },
                ytick style={
                        draw=none
                    },
                x tick label style={
                        anchor=north,
                        align=center,
                        text width=1.2cm,
                        yshift=-1mm,
                        font=\linespread{0.8}\selectfont,
                    },
                extra y ticks={
                        0,
                        44.5
                    },
                extra y tick labels={
                        white-box,
                        black-box
                    },
                extra y tick style={
                        tick align=outside,
                        tick pos=left,
                        tick label style={
                                yshift=146mm,
                                xshift=-2mm,
                                rotate=0,
                            },
                        ticklabel pos=top,
                        rotate=-90,
                    },
                extra x ticks={
                        1,
                        2,
                        3.5,
                        4.5,
                        6
                    },
                extra x tick style={
                        tick align=outside,
                        tick pos=left,
                        tick label style={
                                yshift=5mm,
                            },
                        ticklabel pos=top,
                    },
                extra x tick labels={
                        \numbercircled{1},
                        \numbercircled{2},
                        \numbercircled{3},
                        \numbercircled{4},
                        \numbercircled{5}
                    },
            ]
            \node[
                text width=1em,
                anchor=north,
            ] at (rel axis cs: 0.5, 0.5) {
                \phantomsubcaption
                \label{plot:autovc_subjective_result}
            };
            \addplot[
                ybar,
                pattern=crosshatch,
                pattern color=blue,
            ] plot coordinates {
                    (1, 0.5)
                    (2, 68.5)
                    (3.5, 0.0)
                    (4.5, 54.0)
                    (6, 43.0)
                };
            \addplot[
                ybar,
                pattern=crosshatch dots,
                pattern color=teal,
            ] plot coordinates {
                    (1, 1.5)
                    (2, 19.5)
                    (3.5, 2.5)
                    (4.5, 29.5)
                    (6, 30.0)
                };
            \addplot[
                ybar,
                pattern=north east lines,
                pattern color=orange,
            ] plot coordinates {
                    (1, 8.0)
                    (2, 9.0)
                    (3.5, 12.0)
                    (4.5, 15.0)
                    (6, 23.5)
                };
            \addplot[
                ybar,
                pattern=grid,
                pattern color=red,
            ] plot coordinates {
                    (1, 90.0)
                    (2, 3.0)
                    (3.5, 85.5)
                    (4.5, 1.5)
                    (6, 3.5)
                };
            \legend{
                \raisebox{3pt}{(\RomanNumeralCaps{1}) Different, absolutely sure},
                \raisebox{3pt}{(\RomanNumeralCaps{2}) Different, but not very sure},
                \raisebox{3pt}{(\RomanNumeralCaps{3}) Same, but not very sure},
                \raisebox{3pt}{(\RomanNumeralCaps{4}) Same, absolutely sure},
            }
        \end{groupplot}
    \end{tikzpicture}
    \caption{
        Subjective evaluation results with \textbf{embedding attack} for Chou's ($\epsilon = 0.075$) and \textsc{AutoVC} ($\epsilon = 0.05$).
        On $x$-axis, ``adv. input'' stands for adversarial input and ``adv. output'' stands for adversarial output.
    }
    \label{fig:mos_result}
\end{figure*}
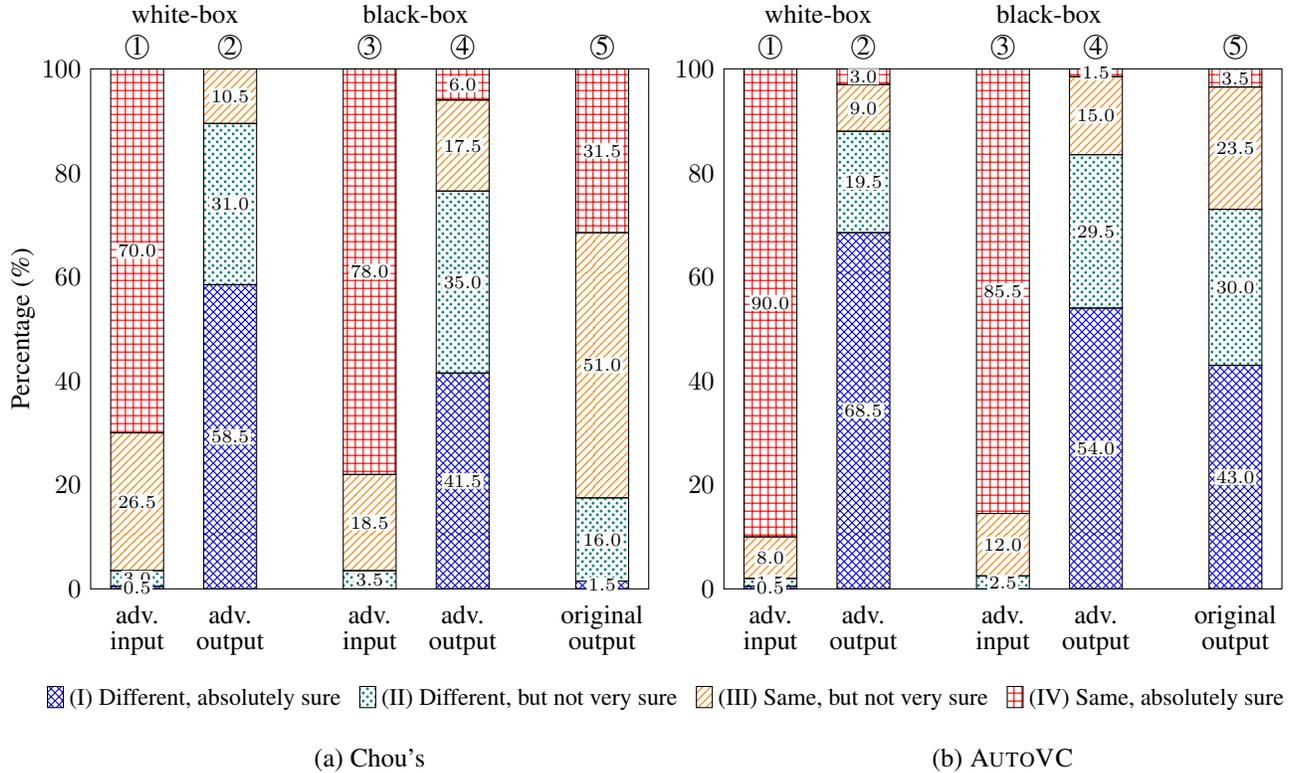

The above speaker verification tests were objective but not necessarily adequate.
So we performed subjective evaluation here but only with the most attractive \textbf{embedding attack} on both Chou's and \textsc{AutoVC} for both \textbf{white-} and \textbf{black-box} scenarios.
We randomly selected 50 out of the 100 example utterances ($\bm{x}$) from the set used in objective evaluation described above.
The corresponding adversarial inputs ($\bm{x} + \bm{\delta}$), outputs ($F (\bm{t}, \bm{x} + \bm{\delta})$) and the original outputs ($F (\bm{t}, \bm{x})$) as used above were then reused in subjective evaluation here for $\epsilon = 0.075$ and $0.05$ respectively for Chou's and \textsc{AutoVC}.
The subjects were then asked to decide if two given utterances were from the same speaker by choosing one out of four: (\RomanNumeralCaps{1}) Different, absolutely sure, (\RomanNumeralCaps{2}) Different, but not very sure, (\RomanNumeralCaps{3}) Same, but not very sure, and (\RomanNumeralCaps{4}) Same, absolutely sure.
Among the two utterances given, one is the original utterance $\bm{x}$, whereas the other is the \textbf{adversarial input}, \textbf{adversarial output}, or \textbf{original output}.
Each utterance pair was evaluated by 6 subjects.
To remove the possible outliers for subjective results, we deleted two extreme ballots on both ends out of the 6 ballots received for each utterance pair (delete a (\RomanNumeralCaps{1}) if there is one, or delete a (\RomanNumeralCaps{2}) if there is no (\RomanNumeralCaps{1})'s, etc.; similar for (\RomanNumeralCaps{4}) and (\RomanNumeralCaps{3})).
In this way 4 ballots were collected for each utterance pair, and 200 ballots for the 50 utterance pairs.
The percentages of ballots choosing (\RomanNumeralCaps{1}), (\RomanNumeralCaps{2}), (\RomanNumeralCaps{3}), (\RomanNumeralCaps{4}) out of these 200 are then shown in the bars \numbercircled{1} \numbercircled{2} for white-box, \numbercircled{3} \numbercircled{4} for black-box scenarios, and \numbercircled{5} for original output in Fig.~\ref{fig:mos_result} for Chou's and \textsc{AutoVC} respectively.

For Chou's, we can see in Fig.~\ref{plot:one_shot_subjective_result} at least 70\% - 78\% of the ballots chose (\RomanNumeralCaps{4}) or considered the adversarial inputs preserved the original speaker characteristics very well (red parts in bars \numbercircled{1} \numbercircled{3}), yet at least 41\% - 58\% of the ballots chose (\RomanNumeralCaps{1}) or considered adversarial outputs obviously from a different speaker (blue parts in bars \numbercircled{2} \numbercircled{4}).
Also for the original output at least 82\% of the ballots considered them close to the original speaker ((\RomanNumeralCaps{3}) plus (\RomanNumeralCaps{4}) in bar \numbercircled{5}).
As in Fig.~\ref{plot:autovc_subjective_result} for \textsc{AutoVC}, at least 85\% - 90\% of the ballots chose (\RomanNumeralCaps{4}) (red parts in bars \numbercircled{1} \numbercircled{3}), yet more than 54\% - 68\% of the ballots chose (\RomanNumeralCaps{1}) (blue parts in bars \numbercircled{2} \numbercircled{4}).
However, only about 27\% of the ballots considered that the original outputs are from the same speaker (red and orange parts in bar \numbercircled{5}).
This is probably because the objective speaker verification system used here didn't match human's perception very well based on which the selected original output with similarity to the original utterance above the threshold may not sound to human subjects as produced by the same speaker.
Also for both two models, the black-box scenario was in general more challenging than the white-box one (lower green and blue parts, \numbercircled{4} v.s. \numbercircled{2}), but the approach is still effective to a good extent.
The demo samples can be found at \url{https://yistlin.github.io/attack-vc-demo}, and the source code is at \url{https://github.com/cyhuang-tw/attack-vc}.

\section{Conclusions}

Improved voice conversion techniques imply higher demand for new technologies to defend personal speaker characteristics.
This paper presents the first known attempt to perform adversarial attack on voice conversion.
Three different approaches are proposed and tested on two state-of-the-art voice conversion models in both objective and subjective evaluation with very encouraging results, including for the black-box scenario closer to real applications.

\section{Acknowledgments}

We are grateful to the authors of \textsc{AutoVC} for offering the complete source code for our experiments and Chou's kind help in model training.
We also thank National Center for High-performance Computing (NCHC) for providing computational and storage resources.

\bibliographystyle{IEEEbib}
\bibliography{main}

\end{document}